%% file: main.tex
\documentclass[sigconf]{acmart}
\AtBeginDocument{%
  }

\setcopyright{acmlicensed}
\copyrightyear{2026}
\acmYear{2026}
\acmDOI{XXXXXXX.XXXXXXX}
\acmConference[SIGMOD '26]{2026 ACM SIGMOD/PODS Conference}{Bengaluru, India}

\input{commands}

\begin{document}

\title{Data Agents: Levels, State of the Art, and Open Problems}

\author{Yuyu Luo$^{\dagger}$}
\affiliation{%
	\institution{HKUST (GZ)}
	\city{Guangzhou}
  \country{China}
}
\email{yuyuluo@hkust-gz.edu.cn}

\author{Guoliang Li}
\affiliation{%
	\institution{\mbox{Tsinghua University}}
	\city{Beijing}
  \country{China}
}
\email{liguoliang@tsinghua.edu.cn}

\author{Ju Fan}
\affiliation{%
	\institution{\mbox{Renmin University of China}}
	\city{Beijing}
  \country{China}
}
\email{fanj@ruc.edu.cn}

\author{Nan Tang}
\affiliation{
	\institution{HKUST (GZ)}
	\city{Guangzhou}
  \country{China}
}
\email{nantang@hkust-gz.edu.cn}

\begin{teaserfigure}
	\begin{center}
		\setcounter{figure}{0}
		\captionsetup{type=figure}
        \vspace{-1.5em}
		\includegraphics[width=0.85\textwidth]{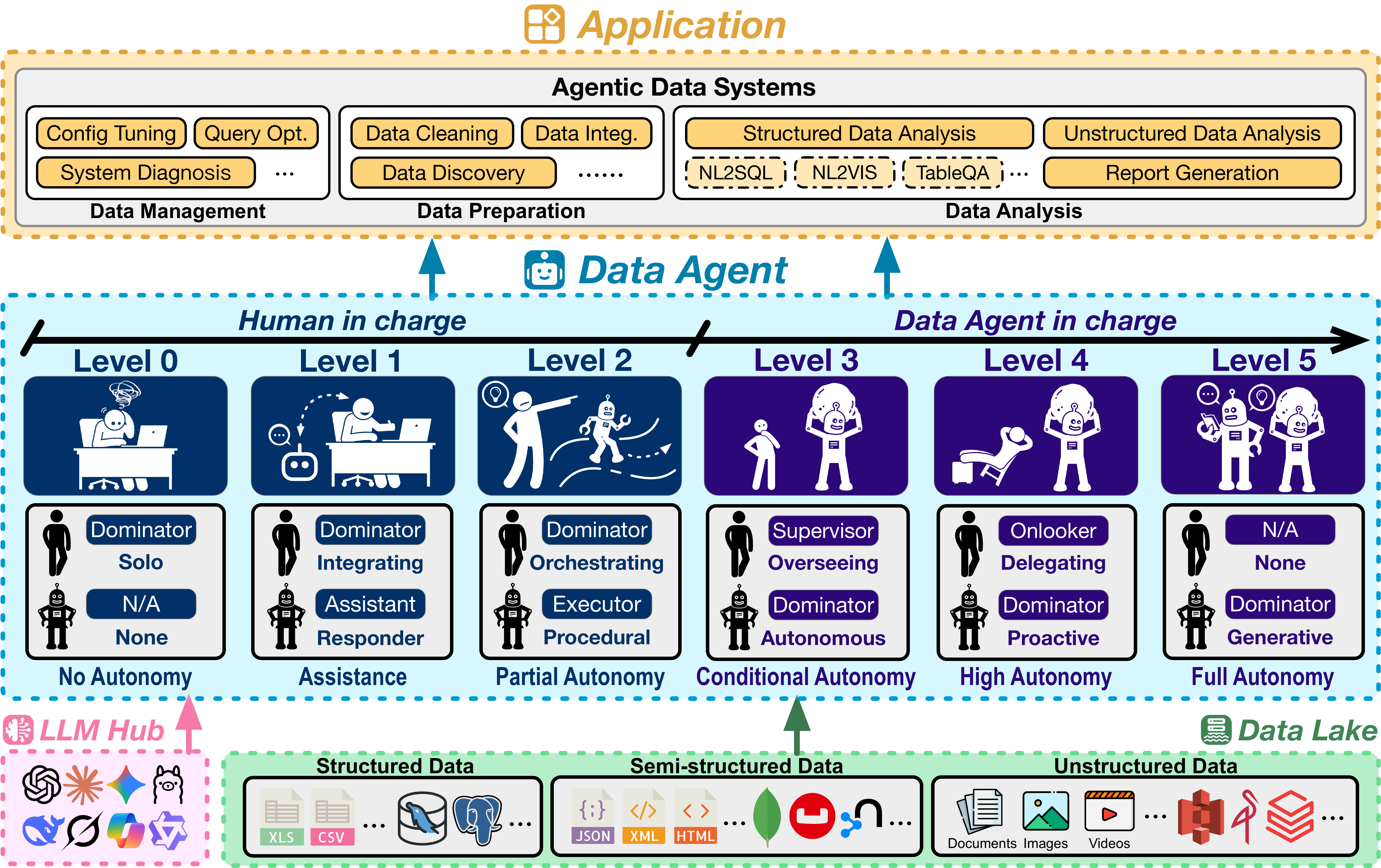}
        \vspace{-.5em}
		\captionof{figure}{\addd{An Overview of Data Agents. (\url{https://github.com/HKUSTDial/awesome-data-agents})}}
        \vspace{-.25em}
		\label{fig:overview}
	\end{center}
\end{teaserfigure}

\input{src/abstract}

\setlist[itemize]{leftmargin=0.4cm, topsep=1pt, itemsep=0pt}
\setlist[enumerate]{leftmargin=0.6cm, topsep=1pt, itemsep=0pt}

\maketitle

\renewcommand{\thefootnote}{\fnsymbol{footnote}}
    \footnotetext[2]{Yuyu Luo is the corresponding author. E-mail: \texttt{yuyuluo@hkust-gz.edu.cn}}
\renewcommand{\thefootnote}{\arabic{footnote}}


\input{src/tutorial}
\input{src/bio}


\newpage
\balance

\bibliographystyle{ACM-Reference-Format}
\bibliography{reference}

\end{document}

%% file: commands.tex
\usepackage{amsmath,amsfonts}
\usepackage{algorithmic}
\usepackage{array}
\usepackage[caption=false,font=normalsize,labelfont=sf,textfont=sf]{subfig}
\usepackage{textcomp}
\usepackage{stfloats}
\usepackage{url}
\usepackage{verbatim}
\usepackage{graphicx}
\usepackage{balance}
\usepackage{amsthm}
\usepackage{color}
\usepackage{xspace}

\usepackage{multirow}
\usepackage{graphicx}
\usepackage{rotating}
\usepackage{array}
\usepackage{pifont}
\usepackage{float}
\usepackage[edges]{forest}
\usepackage{booktabs}
\usepackage{makecell}      
\usepackage{graphicx}     
\usepackage{enumitem}

\definecolor{hidden-draw}{RGB}{20,68,106}
\definecolor{grey}{rgb}{0.5, 0.5, 0.5}

\usepackage[framemethod=tikz]{mdframed}
\usepackage{bbm}

\usepackage{caption}

\usepackage{booktabs}
\usepackage{array}
\renewcommand{\arraystretch}{1.0}

\newcommand{\eat}[1]{}

\definecolor{shadecolor}{RGB}{220,220,220}

\definecolor{inputcolor}{RGB}{255,139,35}
\definecolor{outputcolor}{RGB}{120,212,252}
\definecolor{embedcolor}{RGB}{254,127,156}
\definecolor{maskcolor}{RGB}{122,128,255}
\definecolor{ecolor}{RGB}{58,149,54}

\definecolor{highcolor}{RGB}{255,153,153}
\definecolor{midcolor}{RGB}{255,204,204}
\definecolor{lowcolor}{RGB}{204,229,255}

\definecolor{green}{RGB}{0,128,0}

\definecolor{yellow}{RGB}{255,200,18}

\newcommand{\stab}{\vspace{1.2ex}\noindent}

\newcommand{\bi}{\begin{itemize}}
\newcommand{\ei}{\end{itemize}}

\newcommand{\be}{\begin{enumerate}}
\newcommand{\ee}{\end{enumerate}}
\newcommand{\beqn}{\begin{eqnarray*}}
\newcommand{\eeqn}{\end{eqnarray*}}

\newcommand{\stitle}[1]{\stab\noindent{\bf #1}}

\newcommand{\vs}{\textit{vs.}\xspace}

\newcommand{\eg}{\textit{e.g.,}\xspace}

\newcommand{\greencheck}{{\color{green}\ding{51}}}
\newcommand{\halfcheck}{\color{yellow}\ding{51}\rotatebox[origin=c]{-0.0}{\kern-0.645em\ding{55}}}

\definecolor{github_purple}{rgb}{0.302,0.165,0.498}
\definecolor{hkust_blue}{RGB}{3,48,105}

\NewDocumentCommand{\nan}{ mO{} }{\textcolor{blue}{\textsuperscript{\textit{Nan}}\textsf{\textbf{\small[#1]}}}}

\NewDocumentCommand{\yuyu}{ mO{} }{\textcolor{green}{\textsuperscript{\textit{Yuyu}}\textsf{\textbf{\small[#1]}}}}

\NewDocumentCommand{\xinyu}{ mO{} }{\textcolor{red}{\textsuperscript{\textit{Xinyu}}\textsf{\textbf{\small[#1]}}}}

\NewDocumentCommand{\boyan}{ mO{} }{\textcolor{yellow}{\textsuperscript{\textit{Boyan}}\textsf{\textbf{\small[#1]}}}}

\NewDocumentCommand{\yizhang}{ mO{} }{\textcolor{orange}{\textsuperscript{\textit{Yizhang}}\textsf{\textbf{\small[#1]}}}}

\makeatletter
\apptocmd{\@maketitle}{\centering\insertfig}{}{}
\makeatother

\newcommand{\addd}[1]{\textcolor{black}{#1}}
\renewcommand{\marginpar}[1]{}

\usepackage[skins]{tcolorbox}
\definecolor{colorcommentfg}{HTML}{336633}
\definecolor{colorcommentbg}{HTML}{ededed}
\definecolor{colorcommentframe}{HTML}{336633}

%% file: src/abstract.tex
\begin{abstract}
Data agents are an emerging paradigm that leverages large language models (LLMs) and tool-using agents to automate data management, preparation, and analysis tasks. However, the term ``data agent'' is currently used inconsistently, conflating simple query responsive assistants with aspirational fully autonomous ``data scientists''. This ambiguity blurs capability boundaries and accountability, making it difficult for users, system builders, and regulators to reason about what a ``data agent'' can and cannot do.

In this tutorial, we propose the first hierarchical taxonomy of data agents from Level~0 (L0, no autonomy) to Level~5 (L5, full autonomy). Building on this taxonomy, we will introduce a lifecycle- and level-driven view of data agents. We will (1) present the L0–L5 taxonomy and the key evolutionary leaps that separate simple assistants from truly autonomous data agents, (2) review representative L0–L2 systems across data management, preparation, and analysis, (3) highlight emerging Proto-L3 systems that strive to autonomously orchestrate end-to-end data workflows to tackle diverse and comprehensive data-related tasks under supervision, and (4) discuss forward-looking research challenges towards proactive (L4) and generative (L5) data agents.
We aim to offer both a practical map of today’s systems and a research roadmap for the next decade of data-agent development.

\end{abstract}

%% file: src/tutorial.tex

\section{Introduction}
\label{sec:intro}

Modern data ecosystems are increasingly complex, spanning heterogeneous and multimodal data sources, evolving schemas, and tightly coupled Data+AI pipelines~\cite{li2024llm, zhou2023llm,li2025data+,lead}. At the same time, LLM-based agents have demonstrated strong capabilities in tool use, planning, and iterative reasoning~\cite{liu2025advances, minaee2024large, aot, zhu2024statqa,zhang2024aflow,ncnet2022}. As a result, the term \emph{data agent} has rapidly gained popularity in both academia and industry~\cite{sun2025data, linenet,fu2025autonomous,DBLP:conf/icde/LuoQ0018}, with systems ranging from simple SQL or BI chatbots to ambitious products marketed as fully autonomous ``AI data scientists''.

Without a shared vocabulary, however, fundamentally different systems are being conflated under a single, overloaded term. This leads to mismatched user expectations, ambiguous accountability when failures occur, and difficulty in objectively comparing different approaches. Similar challenges were previously faced by the driving-automation community, which motivated the SAE J3016 standard that introduced a six-level taxonomy of autonomy~\cite{SAE-J3016}.

To address this confusion in data systems community, recent work proposes a hierarchical taxonomy of data agents~\cite{dataagents_survey}, from Level~0 (L0, no autonomy) to Level~5 (L5, full autonomy), together with a structured survey of existing systems along this axis, which describes how task dominance and responsibility gradually shift from human operators to data agents as autonomy increases.

In this tutorial, we build on that survey and turn it into a \emph{teaching-oriented} framework for SIGMOD attendees. Our goal is to help participants (1) understand what different levels of data agents can realistically do, (2) navigate the growing landscape of systems across the data lifecycle, and (3) identify key research challenges for advancing data agents towards higher autonomy.

\vspace{-1em}

\subsection{Tutorial Overview}

We will give a 3-hour tutorial consisting of a 140-minute lecture-style part (Parts~I--IV) followed by a 40-minute \emph{Data Agent Playground} (Part~V) for hands-on exploration and discussion.

\stitle{Part I: Problem Definition and Preliminaries (30 minutes).}
We begin by motivating data agents in modern Data+AI ecosystems and formalizing the notion of a data agent.
We will:
(i) introduce the motivation and problem definition of data agents, emphasizing why existing ``data assistant'' systems are insufficient and why autonomy and responsibility need to be made explicit;
(ii) define data agents more formally and contrast them with general-purpose LLM agents along dimensions such as environment, data scale and structure, error propagation, and governance requirements, using a comparison table to highlight these differences;
(iii) summarize key challenges (terminology ambiguity, lifecycle fragmentation, autonomy vs.\ governance, technical bottlenecks) and motivate the need for a level-based taxonomy of data agents.

\stitle{Part II: L0--L2 Data Agents Across the Data Lifecycle (40 minutes).}
Next, we focus on the lower autonomy levels (L0--L2) and instantiate them in three phases of the data lifecycle: data management, data preparation, and data analysis.
We will:
(i) give an overview of how L0, L1, and L2 manifest in each phase and connect them to the roles of humans and agents illustrated in Figure~\ref{fig:overview};
(ii) deep-dive into each phase: 
in data management, from manual DBAs (L0) to database tuning/diagnosis/query optimization copilots (L1) and L2 agents with direct access to DBMSs and monitoring signals; 
in data preparation, from scripts and rules (L0), to suggestion-style copilots to conduct data cleaning, integration, and discovery (L1), to L2 agents that invoke external tools and close the loop via execution feedback; 
in data analysis, from structured data analysis (Table QA / NL2SQL / NL2VIS), unstructured data analysis, and report generation with prompt-response paradigm (L1) to L2 environment-perceived analysis agents that maintain state and invoke SQL, plotting, and retrieval tools;
(iii) use one or two running examples (e.g., database operations and BI analytics) to make the differences between L0, L1, and L2 concrete.
We conclude this part by summarizing recurring design patterns at L0--L2 and their reliability boundaries.

\stitle{Part III: L3 Data Agents and Proto-L3 Systems (45 minutes).}
We then move to Level~3, the ongoing research frontier where data agents start to act as workflow orchestrators under human supervision.
We will:
(i) formally define L3 and explain the key evolutionary leap from L2 to L3;
(ii) present representative Proto-L3 systems from academia that explore LLM orchestrators, semantic operators, task DAG optimization, and tool evolution to support versatile, cross-task workflows, and discuss their architectures, supported tasks, orchestration strategies, and limitations;
(iii) analyze industrial ``data agent'' products in cloud data platforms and lakehouses, map them onto corresponding levels, and highlight common design patterns (e.g., DAG-based pipeline orchestration, planner–executor separation, multi-agent collaboration mechanism) and current bottlenecks (e.g., predefined operators/tools, limited causal/meta reasoning, constrained task coverage, strong reliance on human-crafted guardrails).

\stitle{Part IV: Towards L4--L5 and Research Roadmap (25 minutes).}
Finally, we complete the lecture part by discussing the visionary Levels~4 and~5 and outlining a research roadmap.
We will:
(i) elaborate the vision of L4 data agents as proactive, long-lived, self-governing components that continuously monitor Data+AI ecosystems, autonomously discover issues and opportunities, and orchestrate pipelines without explicit instructions;
(ii) introduce L5 data agents as generative data scientists that can invent new solutions, algorithms, and paradigms rather than only applying existing methods;
(iii) summarize key open problems, including autonomous orchestration and versatility, causal and meta reasoning, intrinsic motivation and task discovery, long-horizon planning and trade-offs, safety and governance, and benchmarks for autonomy.

\stitle{Part V: Data Agent Playground --- Hands-on Exploration and Discussion (40 minutes).}
The final part is an interactive \emph{Data Agent Playground} that increases audience engagement. We will walk through a few concrete data-agent workflows~\cite{agenticdata,deepanalyze,databricks_dsa,snowflake_cortex,google_bigquery}, show how L1/L2/Proto-L3 agents behave step by step, and invite attendees to try out our own data-agent prototypes. Participants will be encouraged to sketch or refine agents for their own settings, position them on the L0--L5 spectrum, and discuss key trade-offs in autonomy, governance, and reliability, followed by a brief Q\&A that ties these insights back to the research roadmap in Part~IV.

\subsection{Our Scope and Goals}

\stitle{Our Distinction from Existing Tutorials.}
Existing tutorials and surveys on LLMs and data systems typically focus on specific aspects such as LLMs for databases and data analysis~\cite{zhou2025survey,tang2025llm,li2024llm,li2025data+,liu2025survey, luo2025natural}, data management for machine learning~\cite{chai2023datamanagement_survey,fernandes2023data, zhou2023llm}, or general-purpose LLM agents and tool-using systems~\cite{agenticdata}. In contrast, our tutorial is distinguished by three aspects:
\begin{enumerate}
  \item \textbf{Level-based view.} We adopt a \emph{level-based} perspective on data agents (L0--L5) that explicitly links autonomy, capability, and responsibility, making it easier to reason about what a ``data agent'' at each level can and cannot do.
  \item \textbf{Holistic lifecycle perspective.} We take a \emph{holistic lifecycle} view, jointly covering data management, data preparation, and data analysis under a unified data-agent framework, rather than treating individual tasks in isolation.
  \item \textbf{Evolutionary leaps and roadmap.} We emphasize the \emph{evolutionary leaps} between levels, especially the crucial L2$\rightarrow$L3 and L3$\rightarrow$L4 transitions, and present a \emph{research roadmap} towards proactive (L4) and generative (L5) data agents, instead of providing an exhaustive but flat catalogue of systems.
\end{enumerate}

\stitle{Target Audience and Learning Outcomes.}
This tutorial is intended for a broad SIGMOD audience, including researchers in databases, data mining, machine learning, AI agents, and data-centric AI; system developers and practitioners building data platforms, lakehouses, or enterprise data stacks; and students who wish to enter the emerging area of data agents. By the end of the tutorial, participants will be able to use the L0--L5 framework to position existing and future systems, distinguish data agents from general-purpose LLM agents, interpret and calibrate vendor claims about ``data agents'', choose appropriate autonomy levels for their own applications, and reason about key design dimensions such as perception, planning, tools, memory, and governance. We assume familiarity with basic database concepts and LLM terminology; the tutorial itself will be self-contained.

\input{tables/comparison_with_general_agent}

\section{Tutorial Outline}
\label{sec:outline}

We first define what we mean by data agents and situate them in the broader landscape of data systems and LLM agents.

\subsection{Background and Problem Definition}

\subsubsection{Problem Description: What is a Data Agent?}

Informally, a \emph{data agent} is an LLM-based architecture that orchestrates a Data+AI ecosystem to perform data-related tasks such as configuration tuning, data cleaning, integration, exploration, and analysis~\cite{sun2025data,fu2025autonomous,zhou2025survey}. 
Formally, we can define a data agent $\mathcal{A}$ that operates on raw data $\mathcal{D}$ within an environment $\mathcal{E}$ (\eg DBMS, code interpreters, APIs, etc.), utilizing LLMs $\mathcal{M}$, ultimately producing an output $\mathcal{O}$ to tackle the data-related task $\mathcal{T}$, abstractly represented as:
\vspace{-.1em}
\begin{equation*}
\mathcal{A}: (\mathcal{T}, \mathcal{D}, \mathcal{E}, \mathcal{M}) \rightarrow \mathcal{O}.
\end{equation*}
\vspace{-1.5em}

This broad formulation captures a spectrum of systems, from simple assistants that suggest SQL queries to aspirational ``AI data scientists'' that autonomously manage and analyze data.

\subsubsection{Task Landscape and Data Agents vs.\ General LLM Agents}

Data agents operate within modern Data+AI ecosystems that span relational databases, data warehouses and lakehouses, data lakes, ETL/ELT pipelines, BI tools, and ML services. Therefore, data agents must reason over large, heterogeneous, and often schema-rich data lakes without exhaustive ingestion~\cite{li2024llm}; interact with dynamic and noisy data and systems whose behavior changes over time~\cite{li2025data+}; and operate inside multi-stage pipelines where errors can silently propagate and amplify, rather than affecting only a single response.

Compared to general-purpose LLM agents, data agents thus face more constrained yet substantially more demanding environments. They also need to satisfy stringent requirements on reliability, governance, and reproducibility that are less prominent in many generic agent settings. Table~\ref{tab:comparison_general_data_agents} summarizes key differences between data agents and general LLM agents along these dimensions.

\subsubsection{Key Challenges}

These characteristics give rise to several challenges that motivate the need for a principled taxonomy for data agents:
\begin{itemize}
  \item \textbf{Ambiguous terminology and overstated claims.}
  Without a shared vocabulary, systems with very different autonomy levels are all marketed as ``data agents'', leading to hype, confusion, and misaligned expectations~\cite{sun2025data,fu2025autonomous}.
  \item \textbf{Fragmentation across the data lifecycle.}
  Data agents must span data management, data preparation, and data analysis over heterogeneous, multi-modal data lakes~\cite{zhou2025survey, tang2025llm}, yet most existing works focus on individual tasks or stages in isolation, making it difficult to reason about end-to-end capabilities and trade-offs.
  \item \textbf{Autonomy vs.\ governance.}
  As autonomy increases, assigning responsibility, defining safe operating regions, and providing guarantees become both more important and more challenging, especially when data agents can autonomously modify data, configurations, or pipelines.
  \item \textbf{Technical bottlenecks.}
  Advancing to higher autonomy levels requires progress in perception (over large, complex data and systems), long-horizon planning and orchestration, memory and continual adaptation, causal and meta reasoning, and robust interaction with dynamic environments.
\end{itemize}

To bring clarity, we adopt a level-based framework for data agents that explicitly links autonomy, capability, and responsibility.

\begin{figure*}[t!]
  \centering
  \vspace{-1em}
  \includegraphics[width=\linewidth]{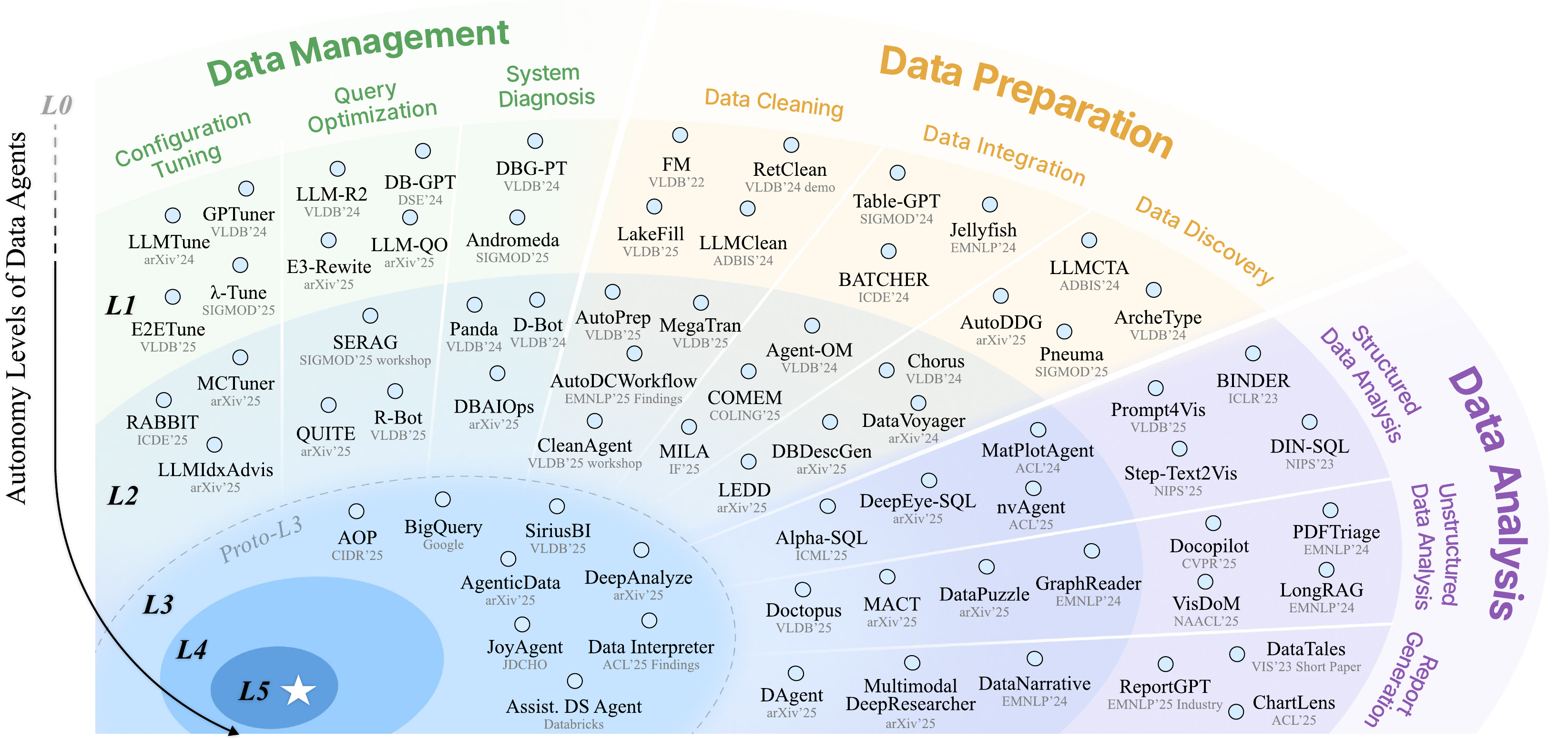}
  \vspace{-2em}
  \caption{Representative Data Agents Across Different Levels.}
  \vspace{-1em}
  \label{fig:representative_work}
\end{figure*}

\subsection{The L0--L5 Hierarchy of Data Agents}
\label{sec:levels}

Inspired by the SAE J3016 standard for driving automation~\cite{SAE-J3016}, we adopt a six-level taxonomy of data agents from L0 to L5. As summarized in Figure~\ref{fig:overview}, data agents are organized into six autonomy levels, from Level~0 (L0) to Level~5 (L5). The figure indicates, for each level, who is in charge of the data-related task (human \vs data agent), what role data agent plays (e.g., responder, executor, orchestrator, proactive or generative component), and which parts of the data lifecycle (management, preparation, analysis) are involved. We briefly review these levels below.

\vspace{1em}
\subsubsection{Levels of Autonomy}

\vspace{-1em}
\paragraph{L0: No Autonomy.}
At L0, there is no data agent involvement. All tasks in data management, preparation, and analysis are performed manually by humans.

\vspace{-.5em}
\paragraph{L1: Assistance.}
L1 data agents operate within a stateless, prompt-response framework. They can answer questions, generate code snippets, or suggest queries, but they do not perceive or interact with the environment. Humans remain fully responsible for executing and verifying any suggestions.

\vspace{-.5em}
\paragraph{L2: Partial Autonomy.}
L2 data agents gain the ability to perceive and interact with their environment, including data lakes, DBMSs, code interpreters, and external APIs. They may possess memory and can invoke tools to autonomously execute task-specific procedures within human-orchestrated pipelines.

\vspace{-.5em}
\paragraph{L3: Conditional Autonomy.}
L3 data agents are expected to autonomously orchestrate and execute tailored data pipelines for a wide range of tasks under human supervision. They interpret high-level user intentions and dominate the end-to-end workflow, while humans act as supervisors.

\vspace{-.5em}
\paragraph{L4: High Autonomy.}
L4 data agents achieve high autonomy and reliability, eliminating the need for human supervision and explicit instructions. They are fully delegated to proactively monitor Data+AI ecosystems, autonomously discover issues and opportunities in data lakes, and orchestrate pipelines to address them.

\vspace{-.5em}
\paragraph{L5: Full Autonomy.}
At L5, data agents are envisioned to innovate new solutions and paradigms beyond existing methods, acting as fully autonomous and generative data scientists. Human involvement becomes unnecessary.

As an overview, Figure~\ref{fig:representative_work} positions representative systems from academia and industry across the L0--L5 levels and the three phases of the data lifecycle.







\subsection{L0--L2: From Manual Workflows to Partial Autonomy}
\label{sec:l0-l2}

In this section, we review representative systems at L0--L2 across three phases of the data lifecycle: data management, data preparation, and data analysis. 

\subsubsection{Data Management}

Data management includes configuration tuning, query optimization, and system diagnosis in database systems~\cite{zhou2023llm, zhao2023automatic}.
At \textbf{L0}, DBAs manually tune knobs, index configurations, and execution plans, relying on expertise and trial-and-error~\cite{zhao2023automatic}.
At \textbf{L1}, LLMs are used as query-responsive assistants to generate tuning suggestions or rewritten queries. They operate in a prompt-response manner, returning recommendations that humans must integrate and validate~\cite{lao2024gptuner, giannakouris2024dbg, li2024llm_r2}. For instance, $\lambda$-Tune~\cite{giannakouris2025lambda} and E2ETune~\cite{huang2025e2etune} use LLMs to recommend configuration candidates based on workload features, and Andromeda~\cite{chen2025andromeda} generates diagnostic suggestions for configuration debugging.
At \textbf{L2}, data agents gain direct access to the DBMS and monitoring information. They can observe workload statistics, execute tuning experiments, and adjust configurations or rewrite queries in a decision loop, while still following human-designed workflows~\cite{yan2025mctuner, zhou2025gaussmaster, song2025quite}. Rabbit~\cite{sun2025rabbit}, R-Bot~\cite{sun2025rbot}, D-Bot~\cite{zhou2024dbot} exemplify this through utilizing environmental feedback in configuration tuning, query rewriting, and system diagnosis.

\subsubsection{Data Preparation}

Data preparation~\cite{fernandes2023data,li2025megatran,chai2023demystifying} covers data cleaning~\cite{llmclean}, integration~\cite{li2024table}, and discovery~\cite{feuer2024archetype}. At \textbf{L1}, data agents primarily act as suggestion engines: RetClean~\cite{RetClean} and LakeFill~\cite{lakefill} infer and impute missing values, LLMClean~\cite{llmclean} generate rules for cleaning tasks, Narayan et al.~\cite{narayan2022can} deploy LLMs to propose schema matches or entity correspondences, AutoDDG~\cite{zhang2025autoddg} and LLMCTA~\cite{korini2025evaluating} produce dataset summaries, metadata, or column annotations. 
Homomorphic compression~\cite{10.1145/3626765} is a promising method for reducing the computational cost of data agents while maintaining semantic integrity.
At \textbf{L2}, data agents go beyond query responder and directly interact with databases or data lakes to execute cleaning and transformation operations, verify constraints, and adjust their strategies based on execution feedback, and iteratively refine integration decisions as more data is explored~\cite{zhang2024sketchfill, qi2025cleanagentautomatingdatastandardization, kayali2023chorus}. Representative systems include CleanAgent~\cite{qi2025cleanagentautomatingdatastandardization}, MegaTran~\cite{li2025megatran} for data cleaning; SEED~\cite{chen2023seed}, Agent-OM~\cite{qiang2024agent} for data integration; LEDD~\cite{an2025ledd} and DBDescGen~\cite{li2025autodb} for data discovery.

\subsubsection{Data Analysis}

Data analysis includes structured and unstructured data analysis, as well as report generation. At \textbf{L1}, we mostly see LLM-driven question-answering assistants for Table QA~\cite{Dater2023, Binder2023, TableMeetsLLM2024}, NL2SQL~\cite{dinsql, supersql, zhu2025elliesql, liu2025nl2sql}, NL2VIS~\cite{Chat2VIS2023IEEE, Prompt4Vis2025VLDB, nvbench2-2025luo,nvBench2021}, textual or multimodal Document QA~\cite{pdftriage2024, docopilot, suri-etal-2025-visdom,xie2024haichart,DBLP:conf/emnlp/WuYSW0L24}, which generate answers to response user questions over curated datasets, and report generators~\cite{cecchi2024reportgpt, sultanum2023datatales, suri-etal-2025-chartlens} that operate on input tables or documents. 
At \textbf{L2}, data agents move beyond static querying to dynamically engage with, verify, and refine multi-step analytical processes~\cite{TableCritic2025,pourreza2025chasesql, shuai2025deepvis, wang2025chartinsighter}. They invoke tools such as SQL engines~\cite{li2025alphasql, li2025deepeyesql}, plotting libraries~\cite{MatPlotAgent2024ACL, nvAgent2025ACL, xu2025dagent}, or retrieval modules~\cite{flare2023, rear2024, zhang2025datapuzzle}, and support iterative exploration and refinement of analyses~\cite{deng2025reforce, Text2Chart312024EMNLP, yang2025deepresearcher}.

\subsection{L3: Striving for Autonomous Data Agents}
\label{sec:l3}

We now turn to Level~3 (L3), which marks a crucial step from procedural executors to autonomous orchestrators.

\subsubsection{From Executor to Dominator}

\begin{figure}[t!]
    \centering
    \vspace{-1em}
    \includegraphics[width=\linewidth]{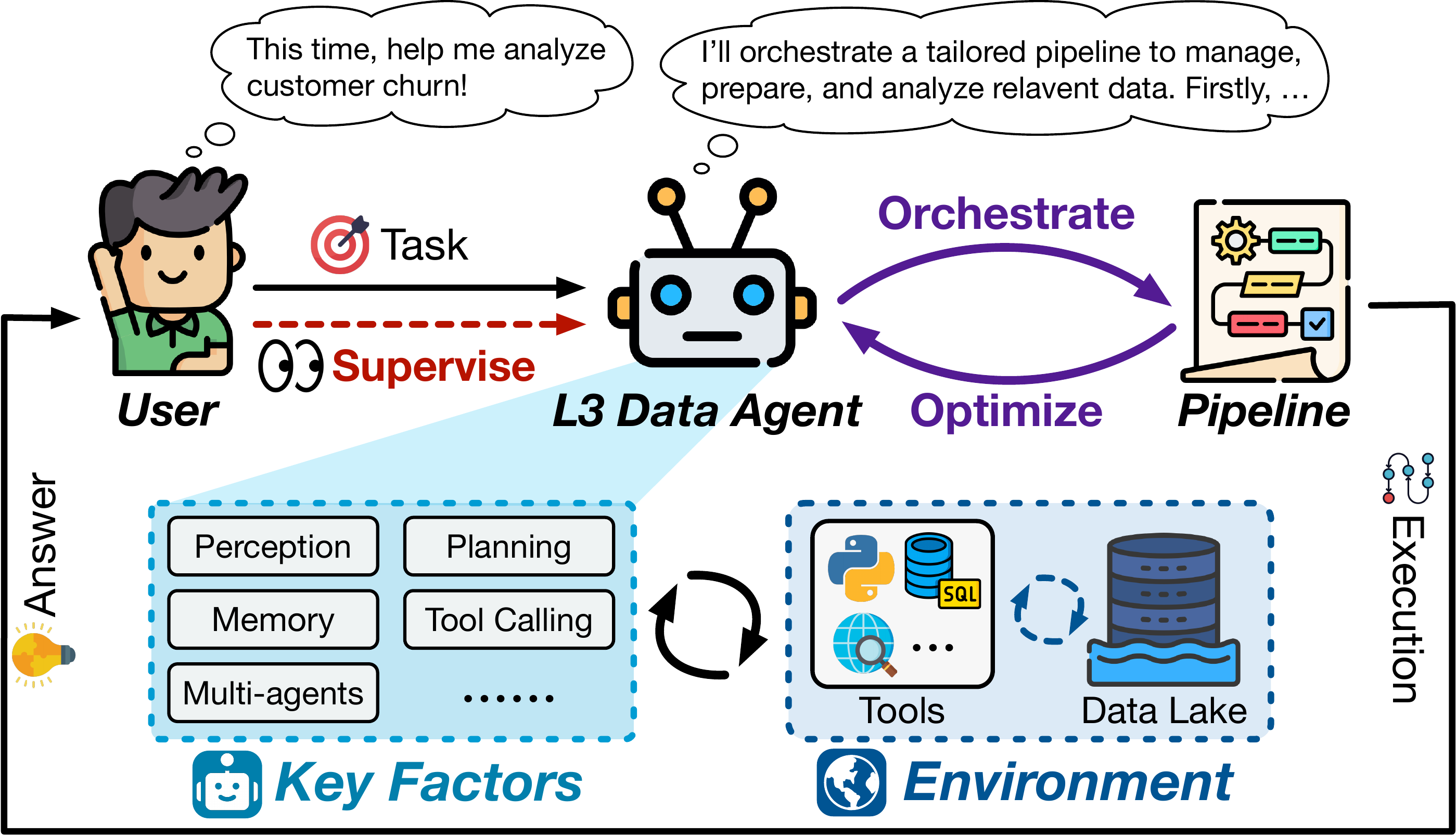}
    \vspace{-1.6em}
    \caption{ L3 Data Agents (Conditional Autonomy).}
    \vspace{-1.75em}
    \label{fig:l3_agent}
\end{figure}

At L2, humans design the overall pipelines, and data agents execute specific procedures within these human-prescribed workflows. At L3, by contrast, data agents are expected to interpret high-level user intent and autonomously orchestrate pipelines that span data management, preparation, and analysis. During execution, data agents adapt the pipeline based on feedback and intermediate results, while humans primarily act as supervisors who review plans and outcomes rather than as pipeline designers. In this sense, task dominance and primary responsibility shift from humans to data agents.
Figure~\ref{fig:l3_agent} illustrates the typical L3 data agent, highlighting its conditional autonomy in autonomous pipeline orchestration and optimization.

\subsubsection{Proto-L3 Data Agents in Research}

Recent research systems begin to exhibit partial L3 capabilities. They use LLM-based orchestrators~\cite{agenticdata}, predefined operators~\cite{wang2025aop, wang2025idatalake}, workflow optimization~\cite{wang2025aop, data-interpreter}, and tool libraries~\cite{deepanalyze} to orchestrate multi-step workflows over heterogeneous systems, cover multiple stages of the data lifecycle within a single agentic process, and maintain state across long-running interactions so that they can refine their plans and correct mistakes over time. These Proto-L3 agents typically operate in constrained environments with curated tools and data, but they provide concrete testbeds for studying the transition from execution-focused L2 agents to orchestration-centered L3 systems.

We will present several representative academic systems and discuss:
(i) their pipeline representation, orchestration, and optimization strategies;
(ii) their architectural choices (single vs.\ multi-agent, central vs.\ decentralized planners);
(iii) their approach to tool abstraction and composition; and
(iv) their strategies for incorporating feedback and handling errors.

Table~\ref{tab:l3_comparison} compares representative Proto-L3 data agents from both academia and industrial products along dimensions such as tool flexibility, data complexity, data lifecycle coverage, and specific management, preparation, and analysis tasks they support.

\subsubsection{Industrial Data-Agent Products}

Industrial platforms (e.g., cloud data warehouses and lakehouses) have started to offer commercial ``data agent'' products~\cite{google_bigquery, snowflake_cortex}. We analyze:
(i) how these products map to the L0--L3 levels in practice;
(ii) which guarantees they provide (e.g., human-in-the-loop confirmation, logging, and rollback);
and (iii) common limitations and design trade-offs.

\subsubsection{Current Bottlenecks and Gaps}

The survey identifies several gaps preventing current systems from realizing full L3 autonomy:
\begin{itemize}
  \item limited pipeline orchestration capabilities and reliance on predefined operators;
  \item inadequate higher-order, causal, and meta-reasoning to diagnose and avoid cascading errors;
  \item difficulty adapting to dynamic environments with changing data and workloads;
  \item heavy reliance on human-crafted reinforcement learning setups for alignment and adoption.
\end{itemize}

These challenges motivate the need for new methods that go beyond straightforward tool-calling LLM agents.

\input{tables/l3_comparison_v2}



\subsection{L4--L5: Vision and Research Roadmap}
\label{sec:l4-l5}

Finally, we discuss the visionary Levels~4 and 5 and outline a research roadmap.

\subsubsection{L4: Proactive, High-Autonomy Data Agents}

At L4, data agents are envisioned as proactive, long-lived, and self-governing components of Data+AI ecosystems. Instead of merely reacting to explicit user requests, an L4 agent continuously monitors data lakes, systems, and models, detects phenomena such as data drift, performance regressions, and schema changes, and identifies opportunities such as beneficial materializations, missing indexes, or promising analytical workflows. It is expected to prioritize among these tasks, design and adapt pipelines to address them without explicit instructions, and operate within reliability, safety, and governance constraints even in the absence of human supervision. Typical scenarios include autonomous detection and mitigation of workload shifts, long-horizon management of indexes and materialized views, and continuous quality assurance for critical data assets. Realizing such capabilities not only raises questions about autonomous orchestration across the full data lifecycle but also calls for mechanisms for intrinsic motivation, task discovery in large data ecosystems, and long-horizon planning that reasons about cumulative cost, latency, and data-quality trade-offs.

\subsubsection{L5: Generative Data Agents}

L5 data agents go beyond deploying existing techniques and are conceived as autonomous, generative data scientists. An L5 data agent is expected to identify gaps in current methods, hypothesize new algorithms or representations when existing approaches are insufficient, design and analyze experiments to test these hypotheses, and iteratively refine its own solutions over time. In this vision, the data agent is not only a user of database and ML systems, but also an active contributor to their evolution. Moving towards L5 requires abstractions that allow data agents to manipulate high-level design choices—such as physical designs, query rewrite strategies, data cleaning policies, or learning procedures—while staying grounded in executable systems, as well as causal and meta reasoning supporting principled diagnosis, comparison, and improvement of alternative designs, even pioneering of innovative solutions, novel theories, and new paradigm.

Although fully realized L4 and L5 data agents remain speculative, articulating these levels helps delineate a research roadmap. In the near term, the most pressing challenges lie in making L2 and Proto-L3 agents more robust, transparent, and governable; in the medium term, progress toward L4 will depend on advances in autonomous orchestration, task discovery, and long-horizon decision making under multi-objective constraints; and in the longer term, movement toward L5 will hinge on integrating causal and meta reasoning with agent-driven experimentation and on developing evaluation methodologies that capture autonomy, adaptability, and safety beyond traditional task-level accuracy.

\subsubsection{Research Opportunities}

The L0--L5 hierarchy suggests several research directions that are closely tied to core data management problems. A central question is how data agents should perceive and act over large, heterogeneous data lakes: which indexes, materialized views, summaries, or learned representations should serve as their ``senses'', how these structures are exposed as tools, and how agents can orchestrate complex pipelines across management, preparation, and analysis while preserving performance, data quality, and governance guarantees. 

A second theme concerns how data agents are trained and evaluated in realistic environments. Here, operational logs, configuration histories, and telemetry can form the basis for constructing training corpora, adapting agent policies over time, and supporting causal and meta reasoning about failures and improvements. This, in turn, calls for benchmarks and methodologies that go beyond task-level accuracy to capture autonomy, robustness, adaptability, and safety on realistic data-management workloads.

%% file: tables/comparison_with_general_agent.tex
\begingroup
\renewcommand{\arraystretch}{1}
\begin{table*}[t!]
\vspace{-1em}
\caption{Comparison between General LLM Agents and Data Agents}
\vspace{-1em}
\label{tab:comparison_general_data_agents}
\resizebox{\textwidth}{!}{%
\begin{tabular}{|c|l|l|}
\hline
\textbf{Aspect} &
  \multicolumn{1}{c|}{\textbf{General LLM Agents}} &
  \multicolumn{1}{c|}{\textbf{Data Agents}} \\ \hline
\begin{tabular}[c]{@{}c@{}}Primary \\[-1pt] Focus\end{tabular} &
  \begin{tabular}[c]{@{}l@{}}Task and Content Centric: \\[-1pt] \textit{Completing defined tasks or generating content.}\end{tabular} &
  \begin{tabular}[c]{@{}l@{}}Data-Lifecycle Centric: \\[-1pt] \textit{Data management, preparation, and analysis.}\end{tabular} \\ \hline
\begin{tabular}[c]{@{}c@{}}Problem \\[-1pt] Scope\end{tabular} &
  \begin{tabular}[c]{@{}l@{}}Self-contained and Static: \\[-1pt] \textit{Acts on explicit instructions and a finite prompt.}\end{tabular} &
  \begin{tabular}[c]{@{}l@{}}Exploratory and Dynamic: \\[-1pt] \textit{Actively explores and navigates vast, dynamic data lakes.}\end{tabular} \\ \hline
\begin{tabular}[c]{@{}c@{}}Input \\[-1pt] Data\end{tabular} &
  \begin{tabular}[c]{@{}l@{}}Small-Scale and Ready-to-Use: \\[-1pt] \textit{Typically receives manageable, clean inputs.}\end{tabular} &
  \begin{tabular}[c]{@{}l@{}}Large-Scale and ``Raw'': \\[-1pt] \textit{Designed to handle heterogeneous, dynamic, and noisy raw data.}\end{tabular} \\ \hline
\begin{tabular}[c]{@{}c@{}}Tool \\[-1pt] Invocation\end{tabular} &
  \begin{tabular}[c]{@{}l@{}}General-Purpose Toolkit: \\[-1pt] \textit{Web search, calculators, OCR, image generators, etc.}\end{tabular} &
  \begin{tabular}[c]{@{}l@{}}Specialized Data Toolkit: \\[-1pt] \textit{DB loaders, SQL equivalence checker, visualization libraries, etc.}\end{tabular} \\[-1pt] \hline
\begin{tabular}[c]{@{}c@{}}Primary \\ Output\end{tabular} &
  \begin{tabular}[c]{@{}l@{}}Generative Artifacts: \\[-1pt] \textit{Human-consumable product: dialogues, reasoning, images, etc.}\end{tabular} &
  \begin{tabular}[c]{@{}l@{}}Data Products and Insights: \\[-1pt] \textit{Config, processed data, insights, visualizations, analytical report, etc.}\end{tabular} \\ \hline
\begin{tabular}[c]{@{}c@{}}Error \\[-1pt] Consequence\end{tabular} &
  \begin{tabular}[c]{@{}l@{}}Localized: \\[-1pt] \textit{Typically affects limited to only the direct output.}\end{tabular} &
  \begin{tabular}[c]{@{}l@{}}Cascading: \\[-1pt] \textit{Errors can cascade, affecting downstream insights.}\end{tabular} \\ \hline
\end{tabular}%
}
\vspace{-0.75em}
\end{table*}
\endgroup

%% file: tables/l3_comparison_v2.tex
\begin{table*}[t!]
\vspace{-1em}
\centering
\caption{\small Comparison of Representative Proto-L3 Data Agents from Academia Research and Industry Products. Compares Open-source: availability; Undef Ops.: capabilities in utilizing unpredefined operators; data-related task coverage across data management, preparation, analysis; data complexity dimensions: Multi-source (Multis.), Heterogeneous (Hete.), and Multimodal (Multim.)}
\vspace{-1em}
\label{tab:l3_comparison}
\resizebox{\textwidth}{!}{%
\begin{tabular}{|c|c|c|c|ccc|ccc|ccc|ccc|}
\hline
\multirow{3}{*}{\textbf{Years}} &
  \multirow{3}{*}{\textbf{Data Agent}} &
  \multirow{3}{*}{\textbf{\begin{tabular}[c]{@{}c@{}}Open-\\ source\end{tabular}}} &
  \multirow{3}{*}{\textbf{\begin{tabular}[c]{@{}c@{}}Undef\\ Ops.\end{tabular}}} &
  \multicolumn{3}{c|}{\textbf{Data Complexity}} &
  \multicolumn{3}{c|}{\textbf{Data Management}} &
  \multicolumn{3}{c|}{\textbf{Data Preparation}} &
  \multicolumn{3}{c|}{\textbf{Data Analysis}} \\ \cline{5-16} 
 &
   &
   &
   &
  \multicolumn{1}{c|}{Multis.} &
  \multicolumn{1}{c|}{Hete.} &
  Multim. &
  \multicolumn{1}{c|}{\begin{tabular}[c]{@{}c@{}}Config \\[-1.5pt] Tun.\end{tabular}} &
  \multicolumn{1}{c|}{\begin{tabular}[c]{@{}c@{}}Query \\[-1.5pt] Opt.\end{tabular}} &
  \begin{tabular}[c]{@{}c@{}}Sys. \\[-1.5pt] Diag.\end{tabular} &
  \multicolumn{1}{c|}{\begin{tabular}[c]{@{}c@{}}Data \\[-1.5pt] Clean.\end{tabular}} &
  \multicolumn{1}{c|}{\begin{tabular}[c]{@{}c@{}}Data \\[-1.5pt] Integ.\end{tabular}} &
  \begin{tabular}[c]{@{}c@{}}Data \\[-1.5pt] Disc.\end{tabular} &
  \multicolumn{1}{c|}{Struct.} &
  \multicolumn{1}{c|}{Unstruct.} &
  \begin{tabular}[c]{@{}c@{}}Report \\[-1.5pt] Gen.\end{tabular} \\ \hline
2025 &
  AgenticData~\cite{agenticdata} &
  - &
  \halfcheck &
  \multicolumn{1}{c|}{\greencheck} &
  \multicolumn{1}{c|}{\greencheck} &
  - &
  \multicolumn{1}{c|}{-} &
  \multicolumn{1}{c|}{\greencheck} &
  \greencheck &
  \multicolumn{1}{c|}{\greencheck} &
  \multicolumn{1}{c|}{\greencheck} &
  \greencheck &
  \multicolumn{1}{c|}{\greencheck} &
  \multicolumn{1}{c|}{\greencheck} &
  - \\ \hline
2025 &
  DeepAnalyze~\cite{deepanalyze} &
  \greencheck &
  - &
  \multicolumn{1}{c|}{\greencheck} &
  \multicolumn{1}{c|}{\greencheck} &
  - &
  \multicolumn{1}{c|}{-} &
  \multicolumn{1}{c|}{-} &
  - &
  \multicolumn{1}{c|}{\greencheck} &
  \multicolumn{1}{c|}{\greencheck} &
  \greencheck &
  \multicolumn{1}{c|}{\greencheck} &
  \multicolumn{1}{c|}{\greencheck} &
  \greencheck \\ \hline
2025 &
  AOP~\cite{wang2025aop} &
  - &
  - &
  \multicolumn{1}{c|}{\greencheck} &
  \multicolumn{1}{c|}{\greencheck} &
  \greencheck &
  \multicolumn{1}{c|}{-} &
  \multicolumn{1}{c|}{\greencheck} &
  \greencheck &
  \multicolumn{1}{c|}{\greencheck} &
  \multicolumn{1}{c|}{\greencheck} &
  \greencheck &
  \multicolumn{1}{c|}{\greencheck} &
  \multicolumn{1}{c|}{\greencheck} &
  - \\ \hline
2025 &
  iDataLake~\cite{wang2025idatalake} &
  \greencheck &
  - &
  \multicolumn{1}{c|}{\greencheck} &
  \multicolumn{1}{c|}{\greencheck} &
  \greencheck &
  \multicolumn{1}{c|}{-} &
  \multicolumn{1}{c|}{\greencheck} &
  - &
  \multicolumn{1}{c|}{\greencheck} &
  \multicolumn{1}{c|}{\greencheck} &
  \greencheck &
  \multicolumn{1}{c|}{\greencheck} &
  \multicolumn{1}{c|}{\greencheck} &
  \greencheck \\ \hline
2024 &
  Data Interpreter~\cite{data-interpreter} &
  \greencheck &
  - &
  \multicolumn{1}{c|}{-} &
  \multicolumn{1}{c|}{\greencheck} &
  \greencheck &
  \multicolumn{1}{c|}{-} &
  \multicolumn{1}{c|}{-} &
  - &
  \multicolumn{1}{c|}{\greencheck} &
  \multicolumn{1}{c|}{-} &
  \greencheck &
  \multicolumn{1}{c|}{\greencheck} &
  \multicolumn{1}{c|}{\greencheck} &
  \greencheck \\ \hline \hline
2025 &
   \begin{tabular}[c]{@{}c@{}}JoyAgent~\cite{JoyAgent-JDGenie}\end{tabular} &
  \halfcheck &
  \halfcheck &
  \multicolumn{1}{c|}{\greencheck} &
  \multicolumn{1}{c|}{\greencheck} &
  - &
  \multicolumn{1}{c|}{-} &
  \multicolumn{1}{c|}{-} &
  \greencheck &
  \multicolumn{1}{c|}{\greencheck} &
  \multicolumn{1}{c|}{\greencheck} &
  \greencheck &
  \multicolumn{1}{c|}{\greencheck} &
  \multicolumn{1}{c|}{\greencheck} &
  \greencheck \\ \hline
2025 &
  \begin{tabular}[c]{@{}c@{}}Assist. DS Agent~\cite{databricks_dsa}\end{tabular} &
  - &
  - &
  \multicolumn{1}{c|}{\greencheck} &
  \multicolumn{1}{c|}{\greencheck} &
  - &
  \multicolumn{1}{c|}{-} &
  \multicolumn{1}{c|}{\greencheck} &
  - &
  \multicolumn{1}{c|}{\greencheck} &
  \multicolumn{1}{c|}{\greencheck} &
  \greencheck &
  \multicolumn{1}{c|}{\greencheck} &
  \multicolumn{1}{c|}{\greencheck} &
  \greencheck \\ \hline
2025 &
  \begin{tabular}[c]{@{}c@{}}TabTab~\cite{tabtab}\end{tabular} &
  - &
  - &
  \multicolumn{1}{c|}{\greencheck} &
  \multicolumn{1}{c|}{\greencheck} &
  - &
  \multicolumn{1}{c|}{-} &
  \multicolumn{1}{c|}{-} &
  - &
  \multicolumn{1}{c|}{\greencheck} &
  \multicolumn{1}{c|}{\greencheck} &
  - &
  \multicolumn{1}{c|}{\greencheck} &
  \multicolumn{1}{c|}{\greencheck} &
  \greencheck \\ \hline
2025 &
  \begin{tabular}[c]{@{}c@{}}ByteDance Data Agent~\cite{bytedance_data_agents}\end{tabular} &
  - &
  - &
  \multicolumn{1}{c|}{\greencheck} &
  \multicolumn{1}{c|}{\greencheck} &
  - &
  \multicolumn{1}{c|}{-} &
  \multicolumn{1}{c|}{-} &
  - &
  \multicolumn{1}{c|}{-} &
  \multicolumn{1}{c|}{\greencheck} &
  - &
  \multicolumn{1}{c|}{\greencheck} &
  \multicolumn{1}{c|}{\greencheck} &
  \greencheck \\ \hline
2025 &
  \begin{tabular}[c]{@{}c@{}}BigQuery~\cite{google_bigquery}\end{tabular} &
  - &
  - &
  \multicolumn{1}{c|}{\greencheck} &
  \multicolumn{1}{c|}{\greencheck} &
  - &
  \multicolumn{1}{c|}{-} &
  \multicolumn{1}{c|}{\greencheck} &
  - &
  \multicolumn{1}{c|}{\greencheck} &
  \multicolumn{1}{c|}{\greencheck} &
  \greencheck &
  \multicolumn{1}{c|}{\greencheck} &
  \multicolumn{1}{c|}{-} &
  - \\ \hline
2025 &
  \begin{tabular}[c]{@{}c@{}}Cortex~\cite{snowflake_cortex}\end{tabular} &
  - &
  - &
  \multicolumn{1}{c|}{\greencheck} &
  \multicolumn{1}{c|}{\greencheck} &
  \greencheck &
  \multicolumn{1}{c|}{-} &
  \multicolumn{1}{c|}{-} &
  - &
  \multicolumn{1}{c|}{\greencheck} &
  \multicolumn{1}{c|}{\greencheck} &
  \greencheck &
  \multicolumn{1}{c|}{\greencheck} &
  \multicolumn{1}{c|}{\greencheck} &
  - \\ \hline
2025 &
  \begin{tabular}[c]{@{}c@{}}Xata Agent~\cite{xata_agent}\end{tabular} &
  - &
  - &
  \multicolumn{1}{c|}{\greencheck} &
  \multicolumn{1}{c|}{\greencheck} &
  - &
  \multicolumn{1}{c|}{\greencheck} &
  \multicolumn{1}{c|}{\greencheck} &
  \greencheck &
  \multicolumn{1}{c|}{-} &
  \multicolumn{1}{c|}{-} &
  \greencheck &
  \multicolumn{1}{c|}{-} &
  \multicolumn{1}{c|}{-} &
  - \\ \hline
2025 &
   \begin{tabular}[c]{@{}c@{}}SiriusBI~\cite{SiriusBI}\end{tabular} &
  - &
  - &
  \multicolumn{1}{c|}{-} &
  \multicolumn{1}{c|}{-} &
  - &
  \multicolumn{1}{c|}{-} &
  \multicolumn{1}{c|}{-} &
  - &
  \multicolumn{1}{c|}{\greencheck} &
  \multicolumn{1}{c|}{-} &
  \greencheck &
  \multicolumn{1}{c|}{\greencheck} &
  \multicolumn{1}{c|}{-} &
  \greencheck \\ \hline
\end{tabular}%
}
\vspace{-0.5em}
\end{table*}

%% file: src/bio.tex
\section{BIOGRAPHY}
\label{sec:bio}

\stitle{Yuyu Luo} is an Assistant Professor at The Hong Kong University of Science and Technology (Guangzhou), with an affiliated position at the HKUST. He received his PhD from Tsinghua University in 2023. His research interests include Data Agents, AI4DB, and Data-centric AI.
He has received the Best-of-SIGMOD 2023 Papers.

\stitle{Guoliang Li} is a full professor in the Department of Computer
Science, Tsinghua University. His research interests mainly include
data cleaning and integration, and machine learning for databases.
He got the VLDB 2017 early research contribution award, TCDE 2014
Early Career Award, VLDB 2023 Industry Best Paper Runner-up,
Best of SIGMOD 2023, SIGMOD 2023 research highlight award,
DASFAA 2023 Best Paper Award, and CIKM 2017 Best Paper Award. 

\stitle{Ju Fan} is a Professor at the DEKE Lab, MOE China, and the School of Information, Renmin University of China. He received his PhD from Tsinghua University in 2013 and received the ACM China Rising Star Award and the 2023 SIGMOD Research Highlight Award. Dr. Fan’s main research interests are AI4DB and database systems.

\stitle{Nan Tang} is an Associate Professor at The Hong Kong University of Science and Technology (Guangzhou), with an affiliated position at the HKUST. He has received the VLDB 2010 Best Paper Award, the 2023 SIGMOD Research Highlight Award, and the Best-of-SIGMOD 2023. His main research interests are AI4DB and data-centric AI.